\documentclass[prd,amsmath,amssymb,superscriptaddress,floatfix,nofootinbib,10pt]{revtex4}
\usepackage{times}
\usepackage{amssymb,amsbsy,amsmath,amsfonts}
\usepackage{graphicx}
\usepackage{float}
\usepackage{color}
\usepackage{morefloats}
\usepackage{rotating}
\usepackage{srcltx}
\usepackage{slashed}
\usepackage{multirow}
\usepackage{verbatim}
\usepackage{hyperref}
\usepackage{tabularx}
\usepackage{bm}
\allowdisplaybreaks[4]

\usepackage{bbding}
\usepackage{threeparttable}

\newcommand{\PreserveBackslash}[1]{\let\temp=\\#1\let\\=\temp}
\newcolumntype{C}[1]{>{\PreserveBackslash\centering}p{#1}}
\newcolumntype{R}[1]{>{\PreserveBackslash\raggedleft}p{#1}}
\newcolumntype{L}[1]{>{\PreserveBackslash\raggedright}p{#1}}

\begin{document}

\title{Contrasting  observables related to the $N^*(1535)$  from the molecular or a genuine structure}

\begin{abstract}
  In this work we compare the predictions for the scattering length and effective range of the channels $K^0 \Sigma^+, K^+ \Sigma^0 , K^+ \Lambda$ and $\eta p$, assuming the $N^*(1535)$ state as a molecular state of these channels, or an original genuine state, made for instance from three quarks. Looking at very different scenarios, what we conclude is that the predictions of these two pictures are drastically different, to the point that we advise the measurement of these magnitudes, accessible for instance by measuring correlation functions, in order to gain much valuable information concerning the nature of this state. 
  
\end{abstract}


\date{\today}

\author{Hai-Peng Li}
\affiliation{Department of Physics, Guangxi Normal University, Guilin 541004, China}

\author{Jing Song}
\email[]{Song-Jing@buaa.edu.cn}
\affiliation{School of Physics, Beihang University, Beijing, 102206, China}
\affiliation{Departamento de Física Teórica and IFIC, Centro Mixto Universidad de Valencia-CSIC Institutos de Investigación de Paterna, 46071 Valencia, Spain}

\author{Wei-Hong Liang}
\email[]{liangwh@gxnu.edu.cn}
\affiliation{Department of Physics, Guangxi Normal University, Guilin 541004, China}
\affiliation{Guangxi Key Laboratory of Nuclear Physics and Technology,
Guangxi Normal University, Guilin 541004, China}

\author{R. Molina}
\email[]{raquel.molina@ific.uv.es}
\affiliation{Department of Physics, Guangxi Normal University, Guilin 541004, China}
\affiliation{Departamento de Física Teórica and IFIC, Centro Mixto Universidad de Valencia-CSIC Institutos de Investigación de Paterna, 46071 Valencia, Spain}

\author{ E. Oset}
\email[]{oset@ific.uv.es}
\affiliation{Department of Physics, Guangxi Normal University, Guilin 541004, China}
\affiliation{Departamento de Física Teórica and IFIC, Centro Mixto Universidad de Valencia-CSIC Institutos de Investigación de Paterna, 46071 Valencia, Spain}
\maketitle

\section{introduction}
The $N^*(1535)~(1/2^{-})$ state is a subject of continuous debate concerning its nature. From the standard three quarks nature it poses a problem since the energy is bigger than the one of the first positive parity excitation of the nucleon, the $N^*(1440)$, although this   state can also have a more complicated structure than just three quarks~\cite{Wang:2023snv}. In fact, some works suggest that the $N^*(1535)$  is a mixture of three quarks and some pentaquark configuration~\cite{Hannelius:2000gu,Zou:2005xy}. A different interpretation for the $N^*(1535)$ structure is provided from the molecular picture. Indeed, the $N^*(1535)$ was studied from this perspective in the pioneering work of the chiral unitary approach~\cite{Kaiser:1995eg,Kaiser:1996js}, where it was interpreted as   a bound state of $K\Lambda, ~K\Sigma$. Subsequently, work in this direction was done in Ref.~\cite{Nieves:2001wt}, fitting
$\pi N $ data up to $2$~GeV, and in Ref.~\cite{Inoue:2001ip} including the $\pi\pi N $ decay
channels. Further work along these lines was done in Ref.~\cite{Bruns:2010sv}
including terms of chiral Lagrangians of second order, beyond the
dominant  Weinberg–Tomozawa contact term used in previous works. All
these works corroborated the claims of Refs.~\cite{Kaiser:1995eg,Kaiser:1996js} about the dynamical
origin of the  $N^*(1535)$ state.
Other works {suggest that the three quarks components also play a role in the build up of this state~\cite{Hyodo:2008xr,Sekihara:2015gvw,Sekihara:2014kya}.} More recently, combining elements of phenomenology and lattice simulation results, the authors of Refs.~\cite{Liu:2015ktc,Abell:2023nex} stress the relevance of the three quark components. The work of Ref.~\cite{Liu:2015ktc}, in addition to the three quark core, includes explicitly the $\eta N$ and $\pi N$ channels, while in Ref.~\cite{Abell:2023nex} the $K\Lambda$ component is added to these two channels. Yet, in neither of Refs.~\cite{Liu:2015ktc,Abell:2023nex} is the $K\Sigma$ channel included. However, in the molecular picture the $K\Sigma$ channel is essential, to the point that if this channel is omitted no bound state is obtained~\cite{Molina:2023jov}.

In the present work, we propose an idea that can shed light on this issue. The starting point can be found in the work of Ref.~\cite{Song:2022yvz}, where it was shown that the knowledge of the binding energy, together with the scattering length and effective range of the relevant channels, provide very valuable information concerning the molecular compositeness of the states. This work, where the range of the interaction is explicitly considered, provides an improved approach to the  original idea of Weinberg~\cite{Weinberg:1965zz}, which has obvious deficiencies, like providing a $pn$ molecular probability of $1.68$ for the deuteron, and a negative effective range for $pn$ scattering when it is actually positive (see  more recent works in this direction in Refs.~\cite{Li:2021cue,Kinugawa:2021ykv,Albaladejo:2022sux}).

The original idea of Ref.~\cite{Song:2022yvz} is extended, and in Ref.~\cite{Dai:2023cyo} it is shown that the combined knowledge of the binding energy of the $T_{cc}$ state and the values of the scattering length and effective range of the $D^0D^{*+}$ and $D^+D^{*0}$ channels lead to the conclusion that the $T_{cc}$ state is a molecular state of these two components.

A different approach is followed in Refs.~\cite{Dai:2023kwv,Song:2023pdq}, where starting from a genuine  state, which necessarily couples to some meson-meson components where it is observed experimentally,   the state develops  molecular components. Yet, the molecular probability can be extremely small, but at the price of developing unrealistic scattering lengths and effective ranges, to the point that with present knowledge of these magnitudes one can conclude again that the $T_{cc}$ is a molecular state~\cite{Dai:2023kwv} and so is the $X(3872)$~\cite{Song:2023pdq}.

In the present work we continue with this line to provide an answer to this question for the $N^*(1535)$, by comparing  the predictions of a genuine state on the scattering length and effective range for the $K^0\Sigma^+$,  $ K^+\Sigma^0 $,  $K^+ \Lambda $, and $ \eta p$ channels, together with the binding energy and width of the state, with the same results obtained with the molecular picture.

At present, there is only information of the $K^+\Lambda$ correlation function~\cite{ALICE:2023wjz,ALICE:2020wvi}. The correlation functions for the other channels
 can be easily accessible by measuring these correlation functions in present facilities as done by the ALICE, CMS or STAR collaborations~\cite{STAR:2014dcy,ALICE:2017jto,STAR:2018uho,ALICE:2018ysd,ALICE:2019hdt,ALICE:2019eol,ALICE:2019buq,ALICE:2019gcn,ALICE:2020mfd,ALICE:2021szj,ALICE:2021cpv,Fabbietti:2020bfg,Bernardes:2022btf},
which have spurred a recent wave of theoretical works on the subject~\cite{Morita:2014kza,Ohnishi:2016elb,Morita:2016auo,Hatsuda:2017uxk,Mihaylov:2018rva,Haidenbauer:2018jvl,Morita:2019rph,Kamiya:2019uiw,Kamiya:2021hdb,Liu:2023uly,Albaladejo:2023pzq,Ikeno:2023ojl,Torres-Rincon:2023qll,Liu:2023wfo}.
In the absence of these data, we shall estimate them from the interaction of $K^0\Sigma^+$,  $ K^+\Sigma^0 $,  $K^+ \Lambda $, and $ \eta p$ channels using the chiral unitary approach~\cite{Molina:2023jov}. 
We shall study to which extend is it possible to obtain all this information for the binding energy and scattering data starting from a genuine state. What we shall see is that it is impossible to attain this task, since the predictions of the molecular picture and those of the genuine state are drastically different. 
With these results we provide a strong motivation to undertake the task of measuring these correlation functions, by means of which a more clear picture of the structure if the $N^*(1535)$ will emerge.

\section{formalism}
\subsection{Determination of scattering lengths and effective ranges}
Following Ref.~\cite{Molina:2023jov}  we take the channels $K^0\Sigma^+$,  $ K^+\Sigma^0 $,  $K^+ \Lambda $, and $ \eta p$ that we label 1, 2, 3, 4 respectively. For the purpose of calculating the $N^*(1535)$ bound state, the $\pi N $ channel, which is very far away, can be ignored~\cite{Molina:2023jov}.
The transition potential $V_{ij}$ between channels, determined by the chiral Lagrangians, is given by,

\begin{align}
    V_{i j}=-\frac{1}{4 f_\pi^2} C_{i j}\left(k^0+k^{' 0}\right), ~~f_\pi=93~ \mathrm{MeV},
\end{align}
with
$$
k^0=\frac{s+m_1^2-M_1^2}{2 \sqrt{s}},~~k^{'^0}=\frac{s+m_2^2-M_2^2}{2 \sqrt{s}},
$$
where, $m_1$ and $ M_1$ are the masses of the meson and  baryon in the initial state, and $m_2$ and $ M_2$ are the masses of the meson and baryon in the final state. The coefficients $C_{ij}$ are given in Table~\ref{Vcij} and the threshold masses for the channels are given in Table~\ref{massofcha}.

\begin{table}[H]
\centering
 \caption{The value of $C_{ij}$ coefficients of  different channels.}\label{Vcij}
\setlength{\tabcolsep}{22pt}
\begin{tabular}{c|cccc}
\hline
\hline
$C_{ij}$ &   $K^0\Sigma^+$ &  $ K^+\Sigma^0 $ &  $K^+ \Lambda $ & $ \eta p$ \\
\hline
$K^0\Sigma^+$  & $1$ & $\sqrt{2}$ & $0$ & $-\sqrt{\frac{3}{2}}$  \\
$ K^+\Sigma^0 $ & & $0$ & $0$ & $-\frac{\sqrt{3}}{2}$ \\
$K^+ \Lambda $ &  & & $0$ & $-\frac{3}{2}$ \\
$ \eta p$ &  & & & $0$ \\
\hline
\hline
   \end{tabular}
\end{table}

\begin{table}[H]
\centering
 \caption{The values of threshold  of  different channels (in units of MeV).}\label{massofcha}
\setlength{\tabcolsep}{22pt}
\begin{tabular}{c|cccc}
\hline
\hline
Channel &   $K^0\Sigma^+$ &  $ K^+\Sigma^0 $ &  $K^+ \Lambda $ & $ \eta P$ \\
\hline
Threshold &1686.98 & 1686.32&1609.36&1486.13 \\
\hline
\hline
   \end{tabular}
\end{table}

The unitary scattering matrix $T_{ij}(T)$ is given by,
\begin{align}
    T=[{1-V G}]^{-1}V,
\end{align}
where the diagonal $G$~matrix, $\mathrm{Diag}(G_i)$, the loop function of the meson and baryon propagators of the intermediate states, is given by 
\begin{align}
    G_i=2 M_i~\int_{|\bar{q}|< q_{\text {max }}} \frac{d^3 q}{(2 \pi)^3}~\frac{w_i(q)+E_i(q)}{2~ w_i(q)~E_i(q)}~~\frac{1}{s-\left[w_i(q)+E_i(q)\right]^2+i \epsilon}
\end{align}
with $\omega_i(q)=\sqrt{\vec{q}~^2+m_i^2}, ~~E_i(q)=\sqrt{\vec{q}~^2+M_i^2}$, with $m_i$ and  $M_i$   the masses of the meson and baryon in the loop, and $q_{\text {max }}=630~\mathrm{MeV}$~\cite{Molina:2023jov}.

The relationship of the $T$ scattering amplitude to the one in Quantum Mechanism, $f^\mathrm{Q M}$, is given by 
\begin{align}
    T=-\frac{8 \pi \sqrt{s}}{2 M}\times~f^\mathrm{Q M}\simeq-\frac{8 \pi \sqrt{s}}{2 M}\times~ \frac{1}{-\frac{1}{a}+\frac{1}{2} r_0 k^2-i k}
\end{align}

with 
\begin{align}
    k=\frac{\lambda^{1 / 2}\left(s, m_i^2, M_i^2\right)}{2 \sqrt{s}}.
\end{align}

For the 1st  channel, we have \\
\begin{align}\label{TTQM}
    \left(T_{11}\right)^{-1}=-\frac{2 M_1}{8 \pi \sqrt{s}}\left(f^{Q M}\right)^{-1}
\simeq -\frac{2 M_1}{8 \pi \sqrt{s}}\left(-\frac{1}{a}+\frac{1}{2} r_0 k^2-i k\right),
\end{align}
from where we obtain
$$
-\frac{1}{a}+\frac{1}{2} r_{0} k^2 \equiv-\frac{8 \pi \sqrt{s}}{2 M_1}\left(T_{11}\right)^{-1}+i k,
$$
and, thus, 
\begin{align}
    -\frac{1}{a} = -\frac{8 \pi \sqrt{s}}{2 M_1}\left(T_{11}\right)^{-1}|_{\sqrt{s}_\mathrm{th,1}},
\end{align}
where $\sqrt{s}_\mathrm{th,1}$ is the threshold mass for channel 1. The effective range for this channel   is given by,
$$
\begin{gathered}
r_{0}=\left[2 \frac{\partial}{\partial k^2}\bigg(-\frac{8 \pi \sqrt{s}}{2 M}\left(T_{11}\right)^{-1}+i k\bigg)\right]_{\sqrt{s}_\mathrm{th,1}}, \\
=\frac{1}{\mu_1} \left[\frac{\partial}{\partial \sqrt{s}}\bigg(-\frac{8 \pi \sqrt{s}}{2 M}\left(T_{11}\right)^{-1}+i k\bigg)\right]_{\sqrt{s}_\mathrm{th,1}},
\end{gathered}
$$
with $\mu_1$ the reduced mass of channel 1. The same formulas can be applied to $T_{22}$, $T_{33}$, and $T_{44}$.

The pole for a possible   state is looked for in the second Riemann sheet, and is obtained by using $G^{(\mathrm{II})}(\sqrt{\mathrm{s}})$ as
$$
G^{(\mathrm{II})}(\sqrt{s})=G(\sqrt{s})+i \frac{2 M}{4 \pi \sqrt{s}} k, 
$$
for channels where $\operatorname{Re} \sqrt{s}>\sqrt{s_{\text {th }}}$.
Once the pole is determined, the width is approximately two times the imaginary part of the pole (its absolute value), while the couplings are obtained as
$$
\begin{gathered}
g_j^2=\lim _{\sqrt{s} \rightarrow \sqrt{s}_R}\left(\sqrt{s}-\sqrt{s}_R\right) T_{j j}, \\
g_i g_j=\lim _{\sqrt{s} \rightarrow \sqrt{s}_R}\left(\sqrt{s}-\sqrt{s}_R\right) T_{i j},
\end{gathered}
$$
where  $\sqrt{s}_R=\Tilde{M}_R+i \Gamma_R / 2$ is the  pole position in the complex plane. The couplings can be calculated using Cauchy's theorem as,

\begin{align}
    g_i^2=\frac{r}{2 \pi } \int_0^{2 \pi} T_{i i}(z(\theta)) e^{i \theta} \mathrm{d} \theta, 
\end{align}

with $$
\begin{aligned}
& z=z_0+r e^{i \theta}, z_0=\Tilde{M}_R+i \Gamma_R / 2. 
\end{aligned}
$$

\subsection{The scattering amplitude from a genuine state}
Let us assume that the $N^*(1535)$ is a preexisting state of mass $\Tilde{M}_R$, for instance, made of 3 quarks. Yet, this state will couple to the $K^0\Sigma^+$,  $ K^+\Sigma^0 $,  $K^+ \Lambda $, and $ \eta p$ channels,  and we write the potential as,
\begin{align}\label{Vc}
    V_{ij}^\mathrm{(c)}=\frac{\Tilde{g}_i\Tilde{g}_j}{\sqrt{s}-{M}_R},
\end{align}
where $M_R$ is the bare mass of the genuine state, and $\Tilde{g}_i$ are the couplings to each channel, unknown for the moment. We will impose that this state has isospin $I=1/2$. With the phase convention for the isospin multiplets $(K^+,K^0),~~(\bar{K}^0,-K^-), ~~(-\Sigma^+,\Sigma^0,\Sigma^-)$, we have 
$$
|K\Sigma, I=1/2; I_3=1/2\rangle= \sqrt{\frac{2}{3}}K^0\Sigma^+ + {\sqrt{\frac{1}{3}}}K^+\Sigma^0.
$$
Let $\Tilde{g}$ be the coupling of the $N^*(1535)$ to the $I=1/2~~K\Sigma$ state, then, 

$$
\Tilde{g_1}=\sqrt{\frac{2}{3}}~\Tilde{g}; ~~ \Tilde{g_2}=\sqrt{\frac{1}{3}}~\Tilde{g}.
$$
The $V^\mathrm{(c)}$ matrix of Eq.~(\ref{Vc}) can be written  as 
\begin{align}
V^\mathrm{(c)}=
\left(
\begin{array}{cccc}
 \frac{2 \Tilde{g}^2}{3} & \frac{\sqrt{2} \Tilde{g}^2}{3} & \sqrt{\frac{2}{3}} \Tilde{g} \Tilde{g_3} & \sqrt{\frac{2}{3}} \Tilde{g} \Tilde{g_4} \\
 \frac{\sqrt{2} \Tilde{g}^2}{3} & \frac{\Tilde{g}^2}{3} & \frac{\Tilde{g} \Tilde{g_3}}{\sqrt{3}} & \frac{\Tilde{g} \Tilde{g_4}}{\sqrt{3}} \\
 \sqrt{\frac{2}{3}} \Tilde{g} \Tilde{g_3} & \frac{\Tilde{g} \Tilde{g_3}}{\sqrt{3}} & \Tilde{g}_3^2 & \Tilde{g_3} \Tilde{g_4} \\
 \sqrt{\frac{2}{3}} \Tilde{g} \Tilde{g_4} & \frac{\Tilde{g} \Tilde{g_4}}{\sqrt{3}} & \Tilde{g_3} \Tilde{g_4} & \Tilde{g}_4^2 \\
\end{array}
\right)
\end{align}
The scattering matrix is then given by 
\begin{align}\label{TVc}
    T^\mathrm{(c)} =[{1-V^\mathrm{(c)} G}]^{-1}V^\mathrm{(c)}.
\end{align}
We proceed then to evaluate $a_i,r_{0,i}$, the pole position and the couplings of the state to the different components, using the formulas in the former subsection.

The  four unknown parameters, $\Tilde{g},~\Tilde{g_3},~\Tilde{g_4}$, and $M_R$  can be determined by fitting the results obtained for the  the pole position and the values of $a_i,~r_{0,i}$ with the genuine state to these obtained in the former subsection.

\section{results}
\subsection{The scattering lengths and effective ranges}

In Table~\ref{nonfit} we list the scattering lengths and effective ranges for different channels   mentioned in the former section. Then we look for poles of the scattering matrix in the second Riemann sheet and  evaluate the couplings of the state obtained to each channel. The results are listed in  Table~\ref{pole_Coup_mio}. These results are in good agreement with Ref.\cite{Molina:2023jov} within the uncertainties.
The errors are taken from an analysis of synthetic correlation functions assuming errors in the data as in  present experiments (see Table III of Ref.~\cite{Molina:2023jov})
\begin{table}[H]
\centering
 \caption{Scattering lengths and effective range parameters for channel $i$ (in units of $\mathrm{fm}$) obtained with the chiral unitary approach. The channels are $K^0 \Sigma^+~\text{(1)},~K^+ \Sigma^0~\text{(2)},~K^+ \Lambda~\text{(3)}$, and $\eta p~\text{(4)}$.}\label{nonfit}
\setlength{\tabcolsep}{6pt}
\begin{tabular}{ccccc}
\hline
\hline
{Channel}& $\mathbf{1}$ & $\mathbf{2}$ & $\mathbf{3}$ & $\mathbf{4}$ \\
\hline
$a_i$ & $\phantom{-}0.42~(\pm 0.04)-i~0.60~(\pm 0.03)$ & $\phantom{-}0.3~(\pm 0.01)-i~0.32~(\pm 0.02)$ & $\phantom{-}0.29~(\pm 0.02)-i~0.20~(\pm 0.04)$ & $-0.72~(\pm 0.013)$ \\
$r_{0,i}$ & $-1.13~(\pm 0.2)-i~2.68~(\pm 0.2)$ & $-5.81~(\pm 1.4)+i~8.91~(\pm 0.5)$ & $-3.13~( \pm 0.3)-i~0.24~(\pm 0.6)$ & $-1.73~(\pm 0.13)$ \\
    \hline
   \hline
   \end{tabular}
\end{table}

\begin{table}[H]
\centering
\caption{The value of pole position (in units of MeV) and the coupling constants to various channels for the pole obtained with the chiral unitary approach. }\label{pole_Coup_mio}
\setlength{\tabcolsep}{22pt}
\begin{tabular}{ccccc}
\hline
\hline
$\sqrt{s}_R$ &  $g_1$ & $g_2$ & $g_3$ & $g_4$ \\
\hline
 $1519.52-i~82.21$ & $3.56-i~0.88 $ & $2.52-i~0.63 $ & $3.27-i~0.18 $ & $-2.43 +i~ 1.33 $ \\
   \hline
   \hline
\end{tabular}
\end{table}
It is worth commenting here on the results for the scattering length of $K^+\Lambda$ (channel 3). Changing the notation of Ref.~\cite{ALICE:2023wjz}($f(q)=[\frac{1}{a}+\frac{1}{2}r_0k^2-ik]^{-1}$) in~\cite{ALICE:2023wjz} and ($[-\frac{1}{a}+\frac{1}{2}r_0k^2-ik]^{-1}$ here), $a$ in Ref.~\cite{ALICE:2023wjz} is 
$$
0.61\pm0.03\pm0.03,~~-i~0.23\pm0.06\pm0.04.
$$
The imaginary part agrees with the one of table~\ref{nonfit}, but the real part is twice as large. In Ref.~\cite{Molina:2023jov} it is discussed that the results of ~\cite{ALICE:2023wjz} are obtained by using the Lednicky-Lyuboshits~\cite{Lednicky:1981su} formula with $K^+\Lambda$ as single channel. This formula misses the important contribution of the coupled channels and this is the reason for this discrepancy.

\subsection{Fitting procedure with the genuine state} 
We follow different procedures to fit these data with the input provided by the $T^\mathrm{(c)}$ matrix of Eq.~(\ref{TVc}).

\subsubsection{Strategy 1}
Here we follow the strategy to fit the four parameters to the scattering data  $a_1,~r_{0,1}$ of the first channel, $K^0\Sigma^+$, without imposing any restriction of the position of the pole. We obtain the parameters,
$$
\Tilde{g}={2.654 }, \quad \Tilde{g_3}={0.169 }, \quad \Tilde{g_4}={1.655 }, \quad M_R=1807.44 ~\mathrm{MeV}
$$
and the pole position
is located in {$\sqrt{s}_R=1679.28-i~26.50 ~\mathrm{MeV}$}.

In principle, since we have four parameters and four data, we should expect to reproduce the values of the four magnitudes. The values obtained are shown in Table~\ref{calu_1}. 
The agreement with $a_1$, $r_{0,1}$ with the values of Table~\ref{nonfit} is reasonable, and also for the scattering length of channel 2, but
there is a huge disagreement of the results of the other magnitudes for channels $2,~3,~4$ with respect to the reference values of Table~\ref{nonfit}, and the pole position has appeared quite shifted in mass and   width.
\begin{table}[H]
\centering
 \caption{Scattering lengths and effective range magnitudes for channel $i$ in strategy 3(in units of $\mathrm{fm}$ ).}\label{calu_1}
\setlength{\tabcolsep}{22pt}
\begin{tabular}{ccccc}
\hline
\hline
{Channel} & 1 & 2 & 3 & 4 \\
 \hline
$a_i$ & $\phantom{-}0.51-i~0.58 $ & $\phantom{-}0.27-i~0.33 $ & $-0.003-i~0.001 $ & $0.12$ \\
$r_{0, i}$ & $-1.11-i~4.86 $ & $-20.56-i~0.55 $ & $-404.73-i~59.32 $ & $-3.49$ \\
    \hline
   \hline
   \end{tabular}
\end{table}

\subsubsection{Strategy 2}

A second fit is done to $a_1$ and $r_{0,1}$ of Table~\ref{nonfit} and $\sqrt{s}_R$ of Table~\ref{pole_Coup_mio}, and the errors of the magnitudes are taken from Table~III and IV of Ref.~\cite{Molina:2023jov}, which we reproduce in Table~\ref{nonfit}. We force the appearance of the pole at the right position by putting a very small error in that magnitude, taking $0.1$~MeV and $0.01$~MeV for the mass and the width, respectively.


In this case, the parameters are determined as,
$$
\Tilde{g}={4.282}, \quad \Tilde{g_3}={0.0186}, \quad \Tilde{g_4}={2.783}, \quad M_R=1837.70~\mathrm{MeV},
$$
and the pole position is located in   $\sqrt{s_p}= 1519.52-i~82.21$~MeV.

The corresponding $\chi^2/\mathrm{d.o.f.}$ is about $\sim174.95$ only from the fitted data. We show the results of the  $a_i$ and $r_{0,i}$ magnitudes in Table~\ref{fit_erro_R}. We see that Re($a_1$) is in fair agreement with the result of Table~\ref{nonfit}, but the imaginary part is not. However, we already find an appreciable discrepancy in the results of $r_{0,1}$, both for the real and imaginary parts, with the real part changing sign. The results for channel $2$, which were not fitted, show bigger discrepancies. Those for channel $4$ are not that bad, but the discrepancies in channel $3$ are remarkable, with a scattering length extremely small and the effective range substantially bigger.


\begin{table}[H]
\centering
 \caption{Scattering lengths and effective range magnitudes for channel $i$ in strategy 1(in units of $\mathrm{fm}$ ).}\label{fit_erro_R}
\setlength{\tabcolsep}{22pt}
\begin{tabular}{ccccc}
\hline
\hline
{Channel} & 1 & 2 & 3 & 4 \\
 \hline
$a_i$ & $0.38-i~0.19$ & $\phantom{-}0.20-i~0.10$ & $1.50*10^{-5}-i~-2.66*10^{-5}$ & $-1.30$ \\
$r_{0,i}$ & $0.15-i~4.88$ & $-18.04-i~0.59$ & $-43165-i~13819$ & $-1.30$ \\
    \hline
   \hline
   \end{tabular}
\end{table}

\subsubsection{Strategy 3}

In this subsection, we  simultaneously fit to the scattering parameters of all the channels, $a_i$ and $r_{0,i}$ with $i=1...4$ within the uncertainties from Ref.~\cite{Molina:2023jov} shown in Table~\ref{nonfit}, and fix the pole position listed in Table~\ref{pole_Coup_mio}.

The parameters are determined as,
$$
\Tilde{g}={4.51}, \quad \Tilde{g_3}={1.80}, \quad \Tilde{g_4}={2.92}, \quad M_R=1900.31~\mathrm{MeV},
$$
and the pole position is located in   $\sqrt{s_p}= 1519.71-i~82.21$~MeV. 

The corresponding $\chi^2/\mathrm{d.o.f.}$ is about $\sim154.3$ only from the fitted data.
The  results obtained with this strategy are shown in Table~\ref{fit_erro_mio_4}.
Compared with Table~\ref{fit_erro_R}, these results are closer to those of Table~\ref{nonfit}. Yet, comparing with Table~\ref{nonfit} we find that, again, Re($r_{0,1}$) has opposite sign to the one in Table~\ref{nonfit}. The results for $r_{0,2}$ are quite different, and those of channel $3$ have improved, relative to those of Table~\ref{fit_erro_R}. However, even if we have conducted a fit to all these magnitudes, the global discrepancies are quite large, to the point of having a  $\chi^2/\mathrm{d.o.f.}=154.3$, which indicates a very bad fit.

\begin{table}[H]
\centering
 \caption{Scattering lengths and effective range magnitudes for channel $i$ in strategy 2 (in units of $\mathrm{fm}$ ).}\label{fit_erro_mio_4}
\setlength{\tabcolsep}{22pt}
\begin{tabular}{ccccc}
\hline
\hline
{Channel} & 1 & 2 & 3 & 4 \\
 \hline
$a_i$ & $0.343-i~0.211$ & $\phantom{-}0.181-i~0.111$ & $0.144-i~0.184$ & $-1.054$ \\
$r_{0,i}$ & $0.349-i~5.069$ & $-17.643-i~0.979$ & $-4.337-i~1.634$ & $-1.291$ \\
    \hline
   \hline
   \end{tabular}
\end{table}

As we can see, it is impossible to get the results obtained from the molecular picture, starting with a structure for the $N^*(1535)$ state as a genuine state. The conclusion that we obtain is that both pictures provide extremely different prediction for the binding, scattering lengths and effective ranges in the different channel. This should be sufficient motivation for measuring these observables, with the promise of an important step forward towards determining the precise nature of the $N^*(1535)$ state.

\section{Conclusion}

In this work we address the problem of comparing predictions for the scattering length and effective range for the channels $K^0 \Sigma^+, K^+ \Sigma^0 , K^+ \Lambda$ and $\eta p$, together with the existence of the $N^*(1535)$ state, which is tied to these channels in the molecular picture for this state. We evaluate all these observables for the molecular picture and compare them with those obtained assuming that the $N^*(1535)$ state has its origin as a compact, three quark state. We make the comparison of the two predictions. In the lack of measurements, we   assume that the observables obtained with the chiral unitary approach are real data and then conduct fits to these data assuming that the $N^*(1535)$  is a genuine state. We take different strategies. In the genuine picture we have 4 free parameters: three couplings of the state to the meson baryon components and the bare mass of the state. The first fit is to the scattering length and effective range of the $K^0 \Sigma^+$ channel, in total 4 data, where we would expect that a perfect agreement should be obtained. The agreement found for these four magnitudes is fair. However,
 the results for the non fitted magnitudes, scattering length and effective range for the other 3 channels plus the position of the pole, differ quite significatively in both pictures, genuine and molecular. 
   Another fit is conducted demanding that the pole appears at the right position and fitting the 4 magnitudes of channel 1. The discrepancies between the two pictures are also huge, and one gets an idea of the discrepancies from the value of $\chi^2/\mathrm{d.o.f.}$, which is of the order of 170. A similar fit is done by fitting all the observables, while demanding that the pole appears at the right position. Once again the discrepancies are very large with $\chi^2/\mathrm{d.o.f.}$ also of the order of 150. 
   In conclusion, what we find is that the two pictures lead to incompatible results for the scattering parameters. An alternative view of these tests is that the knowledge of these magnitudes has much power to differentiate between the two extreme pictures for the $N^*(1535)$. This study should serve as a motivation to measure all these magnitudes which are accessible nowadays by means of the study of correlation functions among others.

\section{acknowledgements  }

This work is partly supported by the National Natural Science Foundation of China (NSFC) under Grants No. 11975083, No. 12365019, No. 12247108, and by the Central Government Guidance Funds for Local Scientific and Technological Development, China (No. Guike ZY22096024),  and the China Postdoctoral Science Foundation under Grant No. 2022M720359. R. M. acknowledges support from the CIDEGENT pro-
gram with Ref. CIDEGENT/2019/015, the Spanish Ministerio de Economia y Competitivi-
dad and European Union (NextGenerationEU/PRTR) by the grant with Ref. CNS2022-
13614. This work is also partly supported by the Spanish Ministerio de Economia y Com-
petitividad (MINECO) and European FEDER funds under Contracts No. FIS2017-84038-
C2-1-P B, PID2020-112777GB-I00, and by Generalitat Valenciana under contract PROME-
TEO/2020/023. This work is
also partly supported by the Spanish Ministerio de Economia y Competitividad (MINECO) and European FEDER
funds under Contracts No. FIS2017-84038-C2-1-P B, PID2020-112777GB-I00, and by Generalitat Valenciana under
contract PROMETEO/2020/023. This project has received funding from the European Union Horizon 2020 research
and innovation programme under the program H2020-INFRAIA-2018-1, grant agreement No. 824093 of the STRONG-2020 project. This research is also supported by the Munich Institute for Astro-, Particle and BioPhysics (MIAPbP)
which is funded by the Deutsche Forschungsgemeinschaft (DFG, German Research Foundation) under Germany’s
Excellence Strategy-EXC-2094 -390783311.

\bibliography{refs.bib}
\end{document}